\documentclass[
 reprint,
amsmath,amssymb,
aps,
prl,
longbibliography,
citeautoscript
]{revtex4-1}

\usepackage[Symbol]{upgreek}
\usepackage{amssymb}%花体字母加粗
\usepackage{mathrsfs}%花体字母
\usepackage{graphicx}% Include figure files
\usepackage{dcolumn}% Align table columns on decimal point
\usepackage{bm}% bold math
\usepackage{graphicx}% Include figure files
\usepackage{dcolumn}% Align table columns on decimal point
\usepackage{bm}% bold math

\usepackage[english]{babel}
\usepackage{hyperref}
\usepackage{times}
\usepackage{amsfonts,amssymb,bm}
\usepackage{amsmath}
\usepackage{epsfig}
\usepackage{mathrsfs}
\usepackage{color,soul}
\usepackage{bbold}
\usepackage{subfig}
\usepackage{pstricks}
\usepackage{mathtools}
\usepackage{textcomp}
\usepackage{braket}
\usepackage{empheq}
\usepackage{amsmath}
\usepackage{amsthm}
\usepackage[babel]{csquotes}
\usepackage[font=small]{caption}
\usepackage[labelfont=bf,labelsep=quad,justification=raggedright]{caption}
\usepackage{epstopdf}

\begin{document}

\preprint{APS/123-QED}

\title{Higher-order spatiotemporal wave packets with Gouy phase dynamics}% Force line breaks with \\
%\thanks{Subtitle: Towards high-dimensional classical entanglement.}%

\author{Wangke Yu$^{1}$, Yijie Shen$^{1,2}$}\email{yijie.shen@ntu.edu.sg}
\affiliation{
	{$^{1}$Centre for Disruptive Photonic Technologies, School of Physical and Mathematical Sciences and The Photonics Institute, Nanyang Technological University, Singapore 637378, Singapore}\\
	{$^{2}$School of Electrical and Electronic Engineering, Nanyang Technological University, Singapore 637378, Singapore}
}

\date{\today}
\begin{abstract}
\noindent 
Spatiotemporal (ST) wave packets refer to a broad class of optical pulses whose spatial and temporal dependence cannot be treated separately. Such space time non-separability can induce exotic physical effects such as non-diffraction, non-transverse waves, and sub or superluminal propagation. Here, a family of ST nonseparable pulses is presented, where a modal order is proposed to extend their spatiotemporal structural complexity, analogous to the spatial higher‑order Gaussian modes. The modal order is strongly coupled to the Gouy phase, which can unveil anomalous spatiotemporal dynamics, including ultrafast cycle‑switching evolution, ST self‑healing, and sub/super‑luminal propagation. We further introduce a stretch parameter that stretches the temporal envelope while keeping the Gouy‑phase coefficient unchanged. This stretch invariance decouples pulse duration from modal order, allowing us to tune the few‑cycle width without shifting temporal‑revival positions or altering the phase/group‑velocity laws. Moreover, an approach to analysing the phase velocity and group velocity of the higher‑order ST modes is proposed to quantitatively characterise the sub/super‑luminal effects. The method is universal for a larger group of complex structured ultrafast pulses, laying the basis for both fundamental physics and advanced applications in ultrafast optics and structured light.
\end{abstract}

\maketitle

%\tableofcontents
\noindent\textbf{Introduction} -- It is an everlasting human dream to manipulate extreme structures in electromagnetic field, whereby of recent attraction are spatiotemporally sculptured light pulses~\cite{shen2022roadmap,forbes2021structured,he2022towards}. In particular, the pursuing of ultrashort few-cycle, even single-cycle, pulses is an ultimate aim of extreme fast and intense energy extraction~\cite{mourou2019nobel,strickland2019nobel}. Also, the generation and steering of exotic structures in few-cycle pulses have recently holden the promise of extending fundamental scientific effects in light-mater interaction~\cite{xiang2020intermolecular,shaltout2019spatiotemporal,langer2016lightwave}, nonlinear physics~\cite{luu2015extreme,krogen2017generation,shen2017gain}, and spin-orbital coupling~\cite{rego2019generation,dorney2019controlling,kondakci2017diffraction}, as well hatched a myriad of novel applications in ultrafast microscopy~\cite{davis2020ultrafast,wang2020coherent,chen2014quantum}, large-capacity communications~\cite{xie2018ultra,qiao2020multi,wan2022divergence,pryamikov2022rising}, particle trapping~\cite{nie2018relativistic,hilz2018isolated,zhang2021plasmonic}, material machining~\cite{kerse2016ablation,penilla2019ultrafast,malinauskas2016ultrafast,ni2017three}, to name a few. On the other hand, structured pulses are always carrying extraordinary propagation properties in contrast to the conventional waves, challenging the fundamental physical laws and refreshing people's worldview. For instance, it was reported that the structured pulse can hold counterintuitive subluminal group velocities in vacuum~\cite{giovannini2013characterization,bareza2016subluminal,bouchard2016observation}, less than the widely endorsed speed of light in vacuum, $c$, a fundamental constant of nature. Soon after, the superluminal propagation of twisted pulse in some special cases was also demonstrated~\cite{lyons2018fast,petrov2019speed}. With the advanced control of spatiotemporal light, more complex and anomalous propagation with arbitrary sub- and super-luminal behavior of the light pulse was realized~\cite{sainte2017controlling,saari2018reexamination,kondakci2019optical,bhaduri2020anomalous}, broadening the frontier of modern physics. See also recent reviews and demonstrations on spatiotemporal (ST) wave packets, metrology, and guided-wave implementations~\cite{liu2024pi,hall2023lpr,alonso2024aplp,liu2024optica_tiptoe,su2025nc,willner2025nanoph}.Therefore, people will never stop exploring more generalized extreme electromagnetic pulses, not only enabling unlimited applications, but also as a fundamental scientific endeavour in itself.

For characterizing optical pulses, there are well-established theories to obtain a family of space-time non-separable solutions of Maxwell's equations. However, recent studies highlighted the much broader class of ST wave packets, where the spatial and temporal dependence cannot be treated separately. As an earlier modelling of ST nonseparable pulses, in 1983, Brittingham proposed the localized solutions of Maxwell’s equations to obtain the focused spatiotemporal wave modes~\cite{brittingham1983focus}. Soon after, Ziolkowski developed such space-time nonseparable solutions to the scalar wave equation with moving complex sources~\cite{ziolkowski1985exact}, and proposed that a superposition of such pulses leads to finite energy pulses termed ``electromagnetic directed-energy pulse trains'' (EDEPT)~\cite{ziolkowski1989localized}. Special cases of Ziolkowski’s solutions were studied by Hellwarth and Nouchi, who found closed-form expressions that describe focused single-cycle finite-energy solutions to the Maxwell’s equations~\cite{hellwarth1996focused}. This family of pulses includes both linearly polarized pulses, termed ``flying pancakes'' (FPs)~\cite{feng1999spatiotemporal}, as well as pulses of toroidal symmetry, termed ``flying doughnuts'' (FDs)~\cite{hellwarth1996focused}. In the last few decades, the classical FP and FD pulses have become the widely-endorsed formations of few-cycle solutions for guiding developments of generations and applications of ultrashort pulses. For instance of the linear polarized FP pulse, the studies of its propagation~\cite{feng1999spatiotemporal,feng1998gouy}, diffraction~\cite{feng2000spatiotemporal,porras2002diffraction}, and cavity oscillation~\cite{feng2000spatiotemporal,feng1999iso} paved the way for developing the advanced femtosecond few-cycle mode-locking laser~\cite{keller2003recent,kartner2004few,sudmeyer2008femtosecond,tilma2015recent}. While, the toroidal structured FD pulses with more exotic spatial electrodynamics, than the FP pulses, have holden more promises of exciting applications particularly in the contexts of nonradiating anapole materials~\cite{baryshnikova2019optical,savinov2019optical}, topological information transfer~\cite{zdagkas2019singularities}, probing ultrafast light-matter interactions~\cite{raybould2016focused}, and toroidal excitations in matter~\cite{kaelberer2010toroidal,papasimakis2016electromagnetic}. It was also recently demonstrated that the FD pulses can be generated by tailored metamaterials which convert traditional few-cycle pulse into a FD pulse~\cite{papasimakis2018pulse,zdagkas2022observation}.

In this paper, we propose a generalized family of ST pulses, as solutions of Maxwell’s equations, extending the few‑cycle FP and FD modes to higher order. The higher‑order ST pulses (HOSTP) exhibit nonseparable spatiotemporal evolution, where a multi‑cycle structure can evolve into a single‑cycle upon propagation, breaking the conventional fixed cycle‑number description. We introduce a stretch parameter, which, when adjusted, elongates or shortens the pulse’s temporal envelope. To characterise this structure, we propose an approach to analyse the ST‑dependent phase velocity and group velocity of HOSTP. We verify that the superluminal phase velocity interprets the cycle‑switching effect in HOSTP, and the group velocity shows both subluminal and superluminal distributions due to the exotic structure of HOSTP, without violating causality in free space. The seminal and counterintuitive propagation dynamics of HOSTP has the potential to extend the frontier of fundamental physics. Our dynamics analysis is universal for characterising the dynamics of a large group of complex structured ultrafast pulses and for extending advanced applications in ultrafast optics and structured light. 

\vspace{0.2cm}
\noindent\textbf{Fundamental theories} -- The focused few-cycle electromagnetic pulses are characterized by the localized finite-energy space-time non-separable solution of Maxwell’s equations, and the EDEPT method is the most endorsed way to obtain the exact solution~\cite{ziolkowski1989localized}. The first step is finding a scalar generating function $f(\mathbf{r},t)$ that satisfies Helmholtz’s wave equation:
\begin{equation}
	\left( \nabla^2-\frac{1}{{{c}^{2}}}\frac{{{\partial }^{2}}}{\partial {{t}^{2}}} \right)f\left( \mathbf{r},t \right)=0,
\end{equation}
where $\mathbf{r}$ is spatial coordinate [that can be Cartesian coordinate $(x,y,z)$ or cylindrical coordinate $(r,\theta,z)$], $t$ is time, $c=1/\sqrt{\varepsilon_0\mu_0}$ is the speed of light, and the $\varepsilon_0$ and $\mu_0$ are the permittivity and permeability of medium. The exact solution of $f(\mathbf{r},t)$ can be given by the modified power spectrum method proposed by Ziolkowski~\cite{ziolkowski1985exact,ziolkowski1989localized}:
\begin{equation}
f(\mathbf{r},t)={{f}_{0}}\frac{{{e}^{-s/{{q}_{3}}}}}{\left( {{q}_{1}}+i\tau  \right){{\left( s+{{q}_{2}} \right)}^{\alpha }}}
\end{equation}
where $f_0$ is a normalized constant, $s=r^2/(q_1+i\tau)-i\sigma$, $\tau=z-ct$, $\sigma=z+ct$, $q_1,q_2,q_3$ are real positive adjustable parameters with
units of length, and the real dimensionless parameter $\alpha$ comes from the finite-energy assumption, which must
satisfy $\alpha\ge1$ in order for the electromagnetic pulse has finite energy. 
Next, in order to get the exact solutions of
Maxwell’s equations with electric and magnetic vector fields, $\mathbf{E}(\mathbf{r},t)$ and $\mathbf{H}(\mathbf{r},t)$, we construct the Hertz potential $\mathbf{\Pi}(\mathbf{r},t)=\mathbf{\hat{n}}f(\mathbf{r},t)$, where $\mathbf{\hat{n}}$ can be arbitrary-direction vector, and then the electromagnetic fields of transverse electric (TE) mode can be completely solved by~\cite{ziolkowski1989localized}:
\begin{align}
& \mathbf{E}(\mathbf{r},t)=-{{\mu }_{0}}\frac{\partial }{\partial t}\bm{\nabla} \times \mathbf{\Pi } \\ 
& \mathbf{H}(\mathbf{r},t)=\bm{\nabla} \times \left(\bm{\nabla}\times \mathbf{\Pi } \right) 
\end{align}

\begin{figure*}[t!]
	\centering
	\includegraphics[width=\linewidth]{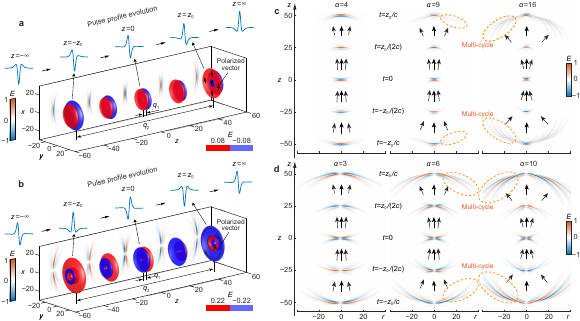}
\caption{\textbf{Spatiotemporal evolutions of fundamental FP and FD pulses:} \textbf{a,b}, The isosurfaces for the electric fields of (\textbf{a}) $E^{\text{(FP)}}(\mathbf{r},t)$ at amplitude levels of $E=\pm0.08$, and (\textbf{b}) $E^{\text{(FD)}}(\mathbf{r},t)$ at amplitude levels of $E=\pm0.22$, with $q_2=100q_1$, at different times of $t=0$, $\pm z_0/(2c)$, and $\pm z_0/c$, inserted with temporal profiles at various spatial positions, $z=0$, $\pm z_0$, and $\pm\infty$, and the $x$-$z$ map plotting the distributions of corresponding instantaneous electric fields. \textbf{Spatiotemporal evolutions of higher-order FP and FD pulses:} \textbf{c,d}, The $r$-$z$ distributions for the electric fields of (\textbf{c}) $E^{\text{(FP)}}_\alpha(\mathbf{r},t)$ with $\alpha=4$, $9$, and $16$, and (\textbf{d}) $E^{\text{(FD)}}_\alpha(\mathbf{r},t)$ with $\alpha=3$, $6$, and $10$, at different times of $t=0$, $\pm z_0/(2c)$, and $\pm z_0/c$, with emerged multi-cycle structures highlighted in dashed lines. The unit for the spatial coordinate is $q_1$. See the detailed dynamic evolutions of various high-order pulses versus index $\alpha$ and time $t$ in Video~1 and Video~2 in Supplementary Materials.} 
	\label{f1}
\end{figure*} 
In the classical few-cycle solutions (FP and FD modes), the finite-energy parameter $\alpha$ is always set as $1$. Here we derive the closed-form expressions for the general finite-energy parameter $\alpha\ge1$. We term it higher-order ST pulses. If the base of the Hertz potential is a unified linear vector $\mathbf{\hat{n}}=\mathbf{\hat{x}}$, we can obtain the closed-form expression of linearly polarized FP pulses dependent on $\alpha$ (polarized along the $y$ axis) (see detailed derivation in Supplementary Information):
\begin{align}
\nonumber
E^{\text{(FP)}}_\alpha(\mathbf{r},t) = &\alpha (\alpha +1){{f}_{0}}\sqrt{\frac{{{\mu }_{0}}}{{{\varepsilon }_{0}}}}{{({{q}_{1}}+i\tau )}^{\alpha -1}}\\
&\times\frac{{{({{q}_{1}}+i\tau )}^{2}}-{{({{q}_{2}}-i\sigma )}^{2}}}{{{\left[ {{r}^{2}}+\left( {{q}_{1}}+i\tau  \right)\left( {{q}_{2}}-i\sigma  \right) \right]}^{\alpha +2}}}
\label{hfp}
\end{align}

For \(\hat{\mathbf n}=\hat{\mathbf z}\) (vortex), we get the $\alpha$-dependent azimuthally polarized electric field as (see detailed derivation in Supplementary Information):
\begin{align}
\nonumber
E^{\text{(FD)}}_\alpha(\mathbf{r},t) = &-\alpha (\alpha +1)i{{f}_{0}}\sqrt{\frac{{{\mu }_{0}}}{{{\varepsilon }_{0}}}}{{({{q}_{1}}+i\tau )}^{\alpha -1}}\\
&\times\frac{r(q_1+q_2-2ict)}{{{\left[ {{r}^{2}}+\left( {{q}_{1}}+i\tau  \right)\left( {{q}_{2}}-i\sigma  \right) \right]}^{\alpha +2}}}
\label{hfd}
\end{align}
where $q_1$ and $z_0=q_2/2$ decide the wavelength and Rayleigh range. 

Under paraxial limit ($q_1\ll q_2$), we can further simplify it into a local-time amplitude-phase expression with clearer manifestation of beam propagation (see detailed derivation in Supplementary Information):
\begin{align}
&E^{\text{(FP)}}_\alpha(\mathbf{r},t)=\frac{\alpha(\alpha+1)w_0^\alpha A_\alpha(\mathbf{r},t)}{2^\alpha z_0^\alpha w^\alpha{{\left( 1+\frac{{{r}^{2}}}{2{w^2}} \right)}^{\alpha+2}}}\text{e}^{i[k_\alpha(\mathbf{r},t )+\alpha\phi(z)]}
\label{hfpa}\\
&E^{\text{(FD)}}_\alpha(\mathbf{r},t)=\frac{i\alpha(\alpha+1)w_0^{\alpha+1} rA_\alpha(\mathbf{r},t)}{2^\alpha z_0^{\alpha+1} w^{\alpha+1}{{\left( 1+\frac{{{r}^{2}}}{2{w^2}} \right)}^{\alpha+2}}}\text{e}^{i[k_\alpha(\mathbf{r},t)+(\alpha+1)\phi(z)]}
\label{hfda}
\end{align}
where, $w(z)=w_0[1+(\frac{z}{z_0})^2]$ and $w_0^2=q_1z_0$ decide the beam radius profile and beam waist size, $\phi(z)=\tan^{-1}(\frac{z}{z_0})$ is the Gouy phase.
The generalized local-time amplitude and wavenumber are defined by $A_\alpha(\mathbf{r},t)=-\frac{f_0\mu_0c(q_1^2+\tau^2)^{(\alpha-1)/2}}{q_1^{\alpha+2}(T^2+1)^{(\alpha+2)/2}}$ and $k_\alpha(\mathbf{r},t)=(\alpha-1)\tan^{-1}(\frac{\tau}{q_1})+(\alpha+2)\tan^{-1}T$. Specially when $\alpha=1$, the higher-order FP and FD modes would be reduced into the fundamental FP and FD pulses~\cite{feng1999spatiotemporal, hellwarth1996focused}.
We note that it is reasonable to name the solutions of $E^{\text{(FP)}}_\alpha(\mathbf{r},t)$ and $E^{\text{(FD)}}_\alpha(\mathbf{r},t)$ as high-order FP and FD pulses, because they have a extremely similar formation as the classical spatial high-order modes, e.g. the high-order Hermite-Gaussian (HG) and Laguerre-Gaussian (LG) modes:
\begin{align}
&E^{\text{(HG)}}_{m,n}(\mathbf{r})=\frac{c_{m,n}}{w}H_m(\widetilde{x})H_n(\widetilde{y})\text{e}^{-\frac{\widetilde{r}^2}{2}}\text{e}^{i[k_\lambda\widetilde{z}+(m+n+1)\phi(z)]}
\label{hg}\\
&E^\text{(LG)}_{p,\ell}(\mathbf{r})=\frac{c_{p,\ell}}{w}\widetilde{r}^{|\ell|}L^{|\ell|}_p(\widetilde{r})\text{e}^{-\frac{\widetilde{r}^2}{2}}\text{e}^{-i\ell\theta}\text{e}^{i[k_\lambda\widetilde{z}+(2p+|\ell|+1)\phi(z)]}
\label{lg}
\end{align}
where $\widetilde{u}=\frac{\sqrt{2}u}{w(z)}$ ($u=x,y,r$), $\widetilde{z}=z+\frac{r^2}{2R(z)}$, $k_\lambda=\frac{2\pi}{\lambda}$ is the wavenumber, $H_m$ and $L_p^\ell$ represent Hermite and generalized Laguerre polynomials, and $c_{m,n}$ or $c_{m,n}$ is the normalized coefficient. Obviously, the order $\alpha$ in high-order FP and FD pulsed modes [Eqs.~(\ref{hfpa}) and (\ref{hfda})] shares a same principal with the order $(m,n)$ and $(p,\ell)$ in HG and LG modes [Eqs.~(\ref{hg}) and (\ref{lg})], that more complicated pattern and multiple Gouy phase are generated with a higher order. Whereas a difference is that, the order in HG and LG modes is limited by nonnegative integer, while the order in high-order FP and FD pulses can be any real number larger than one, i.e. $\alpha\ge1$, due to the basic finite-energy assumption. 

When $\alpha=1$, for both $E^{\text{(FP)}}(\mathbf{r},t)$ and $E^{\text{(FP)}}(\mathbf{r},t)$ modes, the real and imaginary parts can construct two different pulses, i.e. the focused single-cycle pulse and focused $1\frac{1}{2}$-cycle pulse (decided by that the pulse is single-cycle or $1\frac{1}{2}$-cycle at focus), and the single-cycle and $1\frac{1}{2}$-cycle structures can be reshaping upon propagation for both cases.
In contrast to FP mode, the FD has a more complicated amplitude pattern, and the Gouy phase is double that of FP. In this sense, the FD can be seen as a higher-order formation of FP, akin to the high-order spatial mode with complicated pattern and multiple Gouy phase to the fundamental mode~\cite{saleh2019fundamentals}. The Gouy phase is the most crucial factor influencing the temporal evolution in few-cycle pulses~\cite{feng1998gouy}, thus the FP and FD would have different spatiotemporal evolutions, the simulations of which for the focused single-cycle FP and FD pulses are shown in Figs.~\ref{f1}\textbf{a} and \ref{f1}\textbf{b}, respectively. For the FP pulse, a $1\frac{1}{2}$-cycle structure from negative infinity position gradually evolves into the single-cycle at focus, then evolves into another conjugate $1\frac{1}{2}$-cycle structure at positive infinity position. For the FD pulse, the evolution has a stronger space-time non-separability due to the double Gouy phase than the FP pulse, where the negative infinity position shows a single-cycle profile, it already evolves into the $1\frac{1}{2}$-cycle at negative Rayleigh range, next evolves into the conjugate single-cycle at focus, and then to the conjugate $1\frac{1}{2}$-cycle at positive Rayleigh range, finally to the single-cycle at positive infinity position same as the one located at negative infinity. Thus the double Gouy phase in FD reveals the faster temporal reshaping effects. 

\begin{figure*}[t!]
	\centering
	\includegraphics[width=\linewidth]{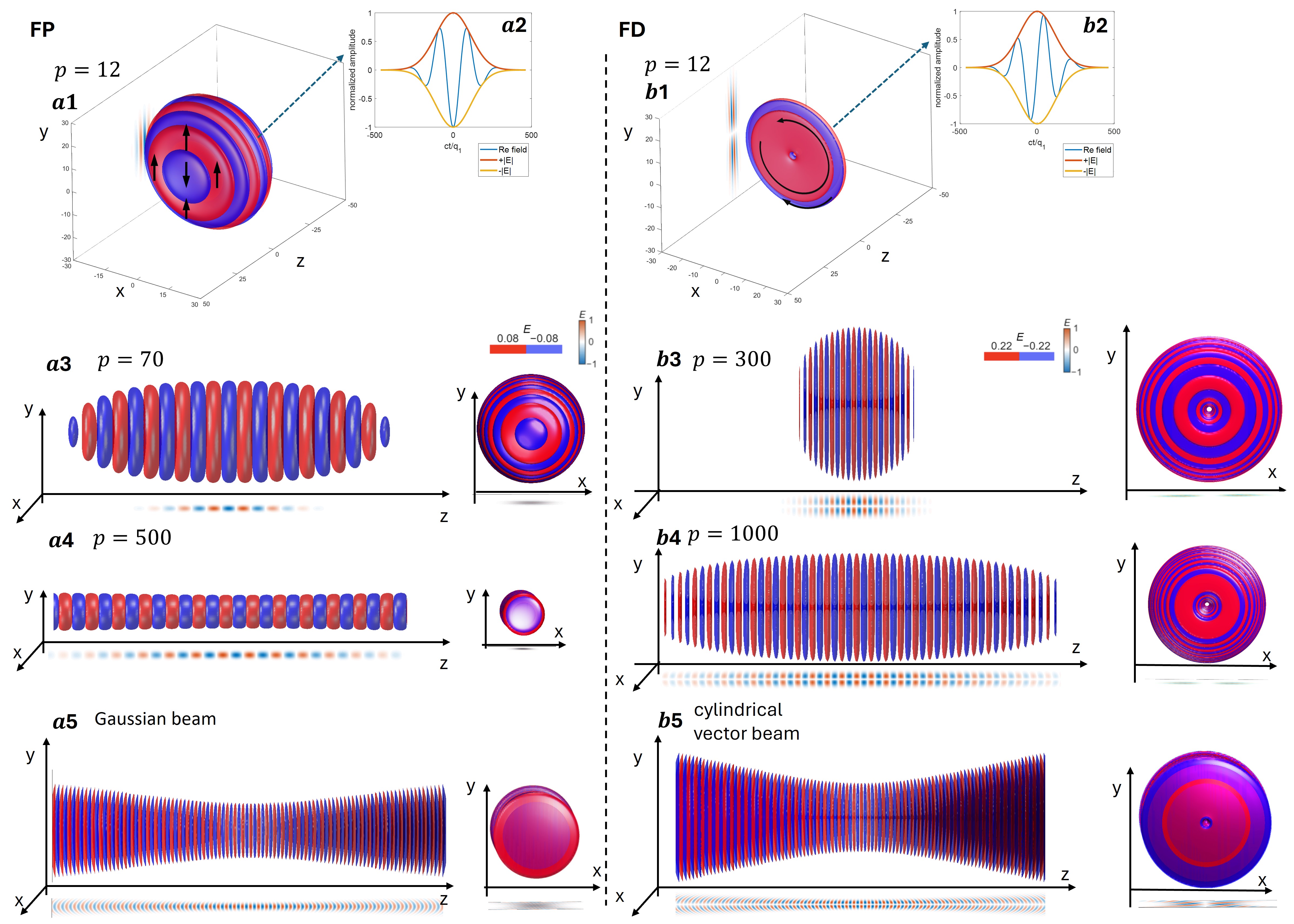}
\caption{\textbf{Iso‑amplitude morphology of FP/FD pulses under the $p$‑stretch and approach to the beam limit.}
Left column: FP; right column: FD. 
\textbf{(a1,b1)} Three‑dimensional Equipotential surface of an electric field for a representative stretch $p=12$; arrows indicate the local polarization (FP: linear $y$; FD: azimuthal). 
\textbf{(a2,b2)} Insets: time traces for the same parameters [instantaneous $\mathrm{Re}\,E$ together with $\pm|E|$]. 
\textbf{(a3,a4)} Three-dimensional isosurfaces of the electric field are shown for FP at representative stretch parameters of $p=70$ and $p=500$. As $p$ increases, a progressively denser stack of half-cycle sheets appears along the local-time direction, leading to an extended temporal envelope.
\textbf{(a5)} In the large‑$p$ limit the FP tends to a linearly polarized Gaussian beam; 
\textbf{(b3,b4)} Three-dimensional isosurfaces of the electric field are shown for FD at representative stretch parameters of $p=300$ and $p=1000$. As $p$ increases, a progressively denser stack of half-cycle sheets appears along the local-time direction, leading to an extended temporal envelope.
\textbf{(b5)} the FD tends to a cylindrical‑vector (azimuthally polarized) beam. 
Red/blue encode positive/negative instantaneous electric field; color bars indicate the normalized iso‑amplitude levels used in the renderings.}

\label{pulse_ReE_envelope}
\end{figure*}

\vspace{0.2cm}
%\begin{figure}[t!]
%	\centering
%	\includegraphics[width=\linewidth]{fs.pdf}
%	\caption{\textbf{Spectral structures of HOSTP:} \textbf{a-h}, The spectral distributions (Fourier transform) at various transverse planes, $z=0,\frac{1}{2}z_0,z_0,\frac{3}{2}z_0$, for (\textbf{a}-\textbf{d}) various high-order FP pulses $|\widetilde{E}^{\text{(FP)}}_\alpha(\mathbf{r},f)|^2$ of $\alpha=1,4,9,16$ and (\textbf{e}-\textbf{h}) high-order FD pulses $|\widetilde{E}^{\text{(FD)}}_\alpha(\mathbf{r},f)|^2$ of $\alpha=1,3,6,10$. The dashed lines mark the parts corresponding to the multi-cycle structures as in Fig.~\ref{f1}.} 	\label{fs}
%\end{figure}
\vspace{0.2cm}
\noindent\textbf{Envelope stretch via parameter \(p\): exact forms and diagnostics} --
Within the EDEPT framework, a spectral phase mask
\[
S_p(a)=\sum_m w_m\,e^{-ia\Delta_m},\qquad \sum_m w_m=1,
\]
acts as a coherent superposition of baseline solutions with a shift \(\sigma\to\sigma-\Delta_m\) for each term. Define
\[
A:=q_1+i\tau,\;\; S_m:=q_2-i(\sigma-\Delta_m),\;\; B_m:=\rho^2 + A\,S_m,\;\; 
\rho\equiv r.
\]
Then the exact FP/FD electric fields with envelope stretch \(p\) are
\begin{align}
E_y^{(\mathrm{FP},p)}(\mathbf r,t)
&= \sum_{m} w_m\,
\Bigl[-\,\alpha(\alpha+1)f_0\sqrt{{\mu_0}/{\varepsilon_0}}\;
A^{\alpha-1}\,
\frac{A^2 - S_m^{2}}{B_m^{\alpha+2}}\Bigr],
\label{eq:EyFPp}\\
E_{\theta}^{(\mathrm{FD},p)}(\mathbf r,t)
&= \sum_{m} w_m\,
\Bigl[+\,\alpha(\alpha+1)i f_0\sqrt{{\mu_0}/{\varepsilon_0}}\;
A^{\alpha-1}\,
\frac{\rho\,(A + S_m)}{B_m^{\alpha+2}}\Bigr].
\label{eq:EthFDp}
\end{align}
The parameter \(p\) controls the length of the temporal envelope through \(\{\Delta_m,w_m\}\) while leaving the Gouy coefficient unchanged (FD exceeds the FP by one unit for the same \(\alpha\)). Figure~\ref{pulse_ReE_envelope} visualizes the $p$-stretch and its beam-limit morphology with iso-amplitude renderings.
For a moderate stretch ($p=12$), panels (a1,b1) show the three-dimensional Equipotential surface of the instantaneous electric field together with the expected polarization distributions (linear $y$ for FP and azimuthal for FD); the insets (a2,b2) provide the corresponding time traces [$\mathrm{Re}\,E$ with $\pm|E|$].
As $p$ increases (FP: a3 $\rightarrow$ a4; FD: b3 $\rightarrow$ b4), the iso-amplitude slices form a progressively denser stack of half-cycle sheets along the local-time direction, i.e., the temporal envelope becomes longer as $p$ increases, while the transverse profile set by $q_1$ and $q_2$ is unchanged and the axial range displayed is kept fixed for comparison.
The panels (a5,b5) highlight the large-$p$ limit: the FP continuously approaches a linearly polarized Gaussian beam, whereas the FD approaches a cylindrical-vector (azimuthally polarized) beam.
These observations are consistent with the $p$-stretch acting on the temporal degree of freedom (envelope lengthening), while leaving the modal polarization and spatial symmetries unchanged.
For detailed derivations, see Supplementary Information.

\noindent\textbf{Spatiotemporal evolution of ST pulses of higher order} -- The spatiotemporal structures for the pulses of higher order FP ($\alpha=4,9,16$) and FD ($\alpha=3,6,10$) are demonstrated in Fig.~\ref{f1}\textbf{c} and \ref{f1}\textbf{d}, respectively, where some basic properties can be observed. For the spatial property, a higher-order pulsed mode is always with a more serious divergence, this principal is same as the conventional higher-order spatial modes. For the temporal profile evolution, higher order pulses always have more rapid temporal remodelling effect; for example, in the 4th-order FP and 3rd-order FD pulses, with four-times Gouy phases, they can undergo twice-cycle structure switching in a single Rayleigh range (it from the single-cycle evolves to the $1\frac{1}{2}$-cycle and then to the conjugate single-cycle at focus). With the order goes even higher, the switch between single- and $1\frac{1}{2}$-cycle structures becomes more frequently, with stronger space-time non-separability. Moreover, we can observe the exotic multi-cycle structures, marked in Fig.~\ref{f1}\textbf{c} and \ref{f1}\textbf{d}, for the very-high-order cases, with a striking pulse evolution effect that: a multi-cycle structure can be evolved into a single-cycle at focus. In addition, with the increasing of order $\alpha$, the HOSTP shows a stronger localization, that the effective energy region is more transversely confined at focus.

\begin{figure}[t]
  \centering
  \includegraphics[width=\linewidth]{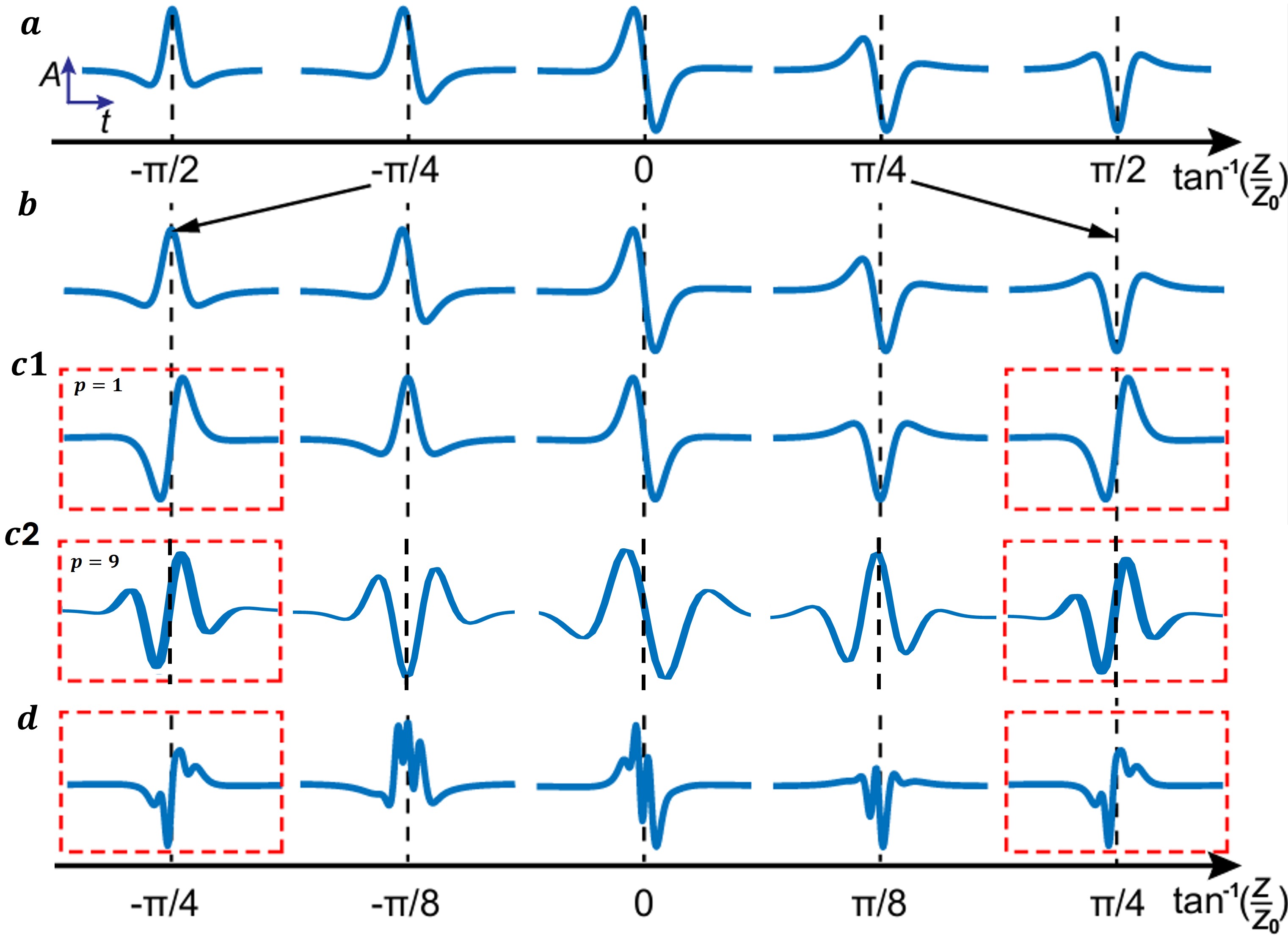} 
  \caption{\textbf{Temporal revivals and \(p\)-stretch invariance.}
  (a) Fundamental FP; (b) fundamental FD; (c) 3rd‑order FD (\(\alpha=3\)); (d) a superposed state \(E^{(\mathrm{FD})}_{\alpha=3}+E^{(\mathrm{FD})}_{\alpha=7}+E^{(\mathrm{FD})}_{\alpha=11}\).
  Vertical dashed lines indicate sampling positions \(\theta=\tan^{-1}(z/z_0)\).
  In (c) we plot two rows labelled \(p=1\) (top) and \(p=9\) (bottom). 
  The red dashed boxes mark the profile‑revival positions at \(\theta=\pm\pi/4\); the boxed waveforms for \(p=1\) and \(p=9\) coincide, demonstrating that while \(p\) stretches the temporal envelope, the Gouy‑phase coefficient and the corresponding revival locations are unaffected by \(p\) (\(p\)-stretch invariance).}
  \label{f2}
\end{figure}

Another intriguing property of spatiotemporal evolution is the structural revival or self-healing that is induced by the fractional Gouy phase effect. For a spatial mode superposed by a series of higher-order eigenmodes, the transverse pattern is usually not preserved upon free propagation, while, it is possible to recover the initial pattern at the specific positions, i.e. a spontaneous self-healing of pattern, due to the fractional Gouy phase effect~\cite{da2020pattern,shen2022self}. Such pattern self-healing effect controlled by Gouy phase has hatched a number of applications from optical trapping to imaging~\cite{da2020pattern,shen2022self}, however, which were never studied in ST domain. Here we demonstrate the ST self-healing effects in HOSTPs. Figures~\ref{f2}\textbf{a} and \ref{f2}\textbf{b} show the temporal profile evolutions of a fundamental FP pulse and FD pulse, same as that inserted in Figs.~\ref{f1}\textbf{a} and \ref{f1}\textbf{b}, respectively, whereby the pulse profile differs through the whole space, thus there is no profile revival effect for both fundamental cases. While, we can observe that the FD has a faster evolution, the pulse profile undertakes the evolution from a $1\frac{1}{2}$-cycle to the opposite $1\frac{1}{2}$-cycle through propagation over the Rayleigh range, but the FP pulse require the propagation throughout the whole space to achieve this. This is exactly induced by the higher coefficient of Gouy phase of FD pulse than that of FP pulse. HOSTPs possess even higher coefficients of Gouy phase dependent on $\alpha$, i.e. $C_\alpha=\alpha$ for the higher-order FP pulses and $C_\alpha=\alpha+1$ for the higher-order FD pulses, which enables the HOSTPs can achieve temporal pattern self-revival or self-healing effect. Figure~\ref{f2}\textbf{c} shows the case of a higher-order FD pulse of $\alpha=3$ (equivalently, a higher-order FD pulse of $\alpha=4$), where the pulse profile can recover to the initial state by propagating over a Rayleigh range. In general, the higher order the pulse is, the faster revival will occur, and which fulfills the relationship of $C_\alpha\tan^{-1}(z/z_0)=2q\pi$ ($q\in\mathbb{Z}$). How many solutions of $q$ corresponds to how many times the revival exists in the whole space. 
You can notice that \ref{f2}\textbf{c} here additionally display two rows labelled $p=1$ and $p=8$ to explicitly test the envelope‑stretch parameter. The equality of the boxed profiles at $\theta=\pm\pi/4$ for $p=1$ and $p=8$ confirms that increasing $p$ only lengthens the temporal envelope (larger FWHM) whereas the Gouy‑phase coefficient $C_\alpha$ (FD larger than FP by one for the same $\alpha$) and the corresponding revival locations $C_\alpha\tan^{-1}(z/z_0)$ remain unchanged.
We can also analyse complex superposed pulses with fractional Gouy phase effect, in this case, the revival repeats only if all Gouy
phases of all pulse components resynchronize along propagation. For instance of the pulse of $E_{\alpha=3}^{\text{(FD)}}+E_{\alpha=7}^{\text{(FD)}}+E_{\alpha=11}^{\text{(FD)}}$, all the component pulses have revival distances, while only at the largest revival distance can all Gouy phases be resynchronized (decided by $E_{\alpha=3}^{\text{(FD)}}$), and the complex pulse profile can revival after propagation of a Rayleigh range (Fig.~\ref{f2}(\textbf{d})).

\begin{figure}[t!]
	\centering
	\includegraphics[width=0.9\linewidth]{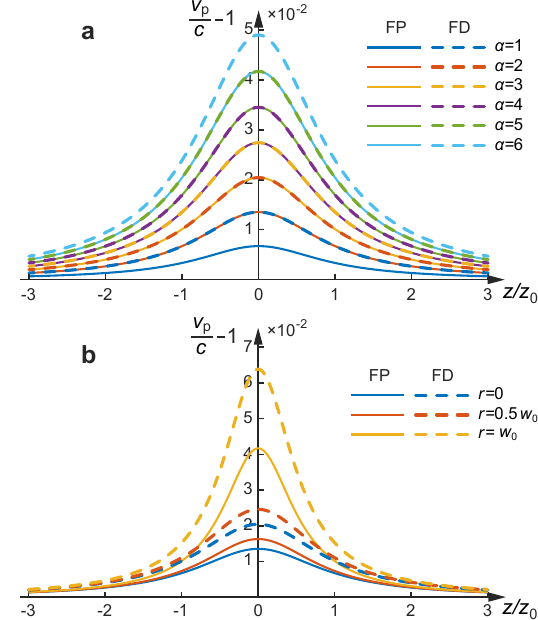}
	\caption{\textbf{Phase velocity of HOSTP:} \textbf{a}, The on-axis phase velocity distributions versus propagation distance of the higher-order FP (solid lines) and FD (dashed lines) pulses with various order indices $\alpha$ from 1 to 6; \textbf{b}, The phase velocity distributions on various off-axis positions from $r=0$ to $r=w_0$ versus propagation distance of the 2nd-order ($\alpha=2$) FP (solid lines) and FD (dashed lines) pulses.} 
	\label{f3}
\end{figure}

\vspace{0.2cm}
\noindent\textbf{Phase velocity} -- The HOSTP has more complex spatiotemporal structure than conventional pulse so that the amplitude envelope and phase are both space-time non-separable expression, i.e. $E(\mathbf{r},t)=A(\mathbf{r},t)\text{e}^{i\varphi(\mathbf{r},t)}$. We evaluate the phase velocity by the original definition that the speed of the traveling of isophase surface, while the phase should be space-time-dependent, so that the conventional calculation of phase velocity would be invalid. Based on Eqs.~(\ref{hfpa}) and (\ref{hfda}), we can find the space-time-dependent phase expressions of the HOSTP as:
\begin{align}
&\varphi^{\text{(FP)}}_\alpha(\mathbf{r},t)=i[k_\alpha(\mathbf{r},t )+\alpha\phi(z)]\\
&\varphi^{\text{(FD)}}_\alpha(\mathbf{r},t)=i[k_\alpha(\mathbf{r},t)+(\alpha+1)\phi(z)]
\end{align}
Then, the isophase surface can be determined by the equation of $\varphi(\mathbf{r},t)=C$ where $C$ is an arbitrary constant. Next, we make differential of the isophase surface equation $\delta\varphi(\mathbf{r},t)=0$ to the variables $z$ and $t$, and the local velocity of the traveling of isophase surface can be derived by $v_\text{p}(r,z)=\delta z/\delta t$, the results of which for the HOSTP are evaluated as (see detailed derivation in Supplementary Information~E): 
\begin{align}
&v_{\text{p}}^{\text{(FP)}}(r,z)=\frac{c}{1-{\frac{\alpha}{z_0}\frac{1}{1+{{\left( {z}/z_0\right)}^{2}}}}/{\left[\frac{3}{q_1}-\frac{(\alpha+2)r^2}{2q_1w^2}\right]}}\label{vpp}\\
&v_{\text{p}}^{\text{(FD)}}(r,z)=\frac{c}{1-{\frac{\alpha +1}{z_0}\frac{1}{1+{{\left( {z}/z_0\right)}^{2}}}}/{\left[\frac{3}{q_1}-\frac{(\alpha+2)r^2}{2q_1w^2}\right]}}\label{vpd}
\end{align}
From Eqs.~(\ref{vpp}) and (\ref{vpd}), the only difference between the two phase velocities is on the coefficients of the denominator terms, $\alpha$ and $\alpha+1$ for the FD and FD pulses, that is directly induced by the different Gouy phase $\alpha\phi(z)$ and $(\alpha+1)\phi(z)$, respectively. Thus the $(\alpha+1)$-order FP pulse has a same phase velocity distribution as the $\alpha$-order FP pulse. That well explain the 4th-order FP and 3rd-order FD pulses shown in the Figs.~\ref{f1}\textbf{c} and \ref{f1}\textbf{d} has the similar cycle-switching evolution (i.e. they both evolve from a single-cycle to $1\frac{1}{2}$-cycle and then to single-cycle in the propagation from position of Rayleigh range to the focus). Another conclusion is that the phase velocity is always superluminal ($v_{\text{p}}>c$), which is consistent with the cycle-switching phenomena that keeps a forward direction as observed above. The on-axis ($r=0$) phase velocity distributions upon propagation of various-order FP nd FD pulses are plotted in Fig.~\ref{f3}\textbf{a}, in a form of $v_{\text{p}}/c-1$ for revealing the difference with speed of light in vacuum. It demonstrates that the phase velocity always meets the maximum at focus ($z=0$) and is decreasing with the propagation position away from the focus. And the phase velocity is always increasing with the increasing of order index $\alpha$. Due to the complex transverse structure of HOSTP, it is meaningful to study the off-axis phase velocity, the phase velocity distributions on various off-axis displacements (from $r=0$ to $r=w_0$) of the 2nd-order ($\alpha=2$) FP and FD pulses are shown in Fig.~\ref{f3}\textbf{b}. It reveals that the phase velocity is increasing at a location with increasing off-axis displacement, which interprets why there is more rapid and complex cycle-switching structure at the off-axis wing regions of a HOSTP as observed hereinbefore.

\vspace{0.2cm}
\noindent\textbf{Group velocity} --
The group velocity is important to discuss because it reveals the actual speed of energy transfer. There are various kinds of definition and formula for evaluating group velocity of structured pulse based on different perspectives~\cite{giovannini2013characterization,bareza2016subluminal,bouchard2016observation,lyons2018fast,petrov2019speed,sainte2017controlling,saari2018reexamination,kondakci2019optical}, however, they are hard to applied in our HOSTP due to its more complex space-time non-separable structure than prior pulses. Here we retrospect the original definition of group velocity that the speed of the traveling of amplitude envelope of a pulse, while there is still a debate on how to evaluate the traveling of amplitude envelope. Our method is firstly finding the centroid surface of the amplitude envelope and then solving the traveling speed of the centroid surface. The expression of the space-time-dependent amplitude envelope of HOSTP can be obtained after simplification (see detailed derivation in Supplementary Information):
\begin{align}
& A_{\alpha }^{\text{(FP)}}(\mathbf{r},t)=\frac{{{\left[ q_{1}^{2}+{{(z-ct)}^{2}} \right]}^{(\alpha -1)/2}}}{{{w}^{\alpha }}{{\left[ {{\left( z-ct+\frac{{{r}^{2}}}{2R} \right)}^{2}}+q_{1}^{\text{2}}{{\left( 1+\frac{{{r}^{2}}}{2{{w}^{2}}} \right)}^{2}} \right]}^{(\alpha +2)/2}}} \label{vgp}\\ 
& A_{\alpha }^{\text{(FD)}}(\mathbf{r},t)=\frac{r{{\left[ q_{1}^{2}+{{(z-ct)}^{2}} \right]}^{(\alpha -1)/2}}}{{{w}^{\alpha +1}}{{\left[ {{\left( z-ct+\frac{{{r}^{2}}}{2R} \right)}^{2}}+q_{1}^{\text{2}}{{\left( 1+\frac{{{r}^{2}}}{2{{w}^{2}}} \right)}^{2}} \right]}^{(\alpha +2)/2}}} \label{vgd}
\end{align}
Based on Eqs.~(\ref{vgp}) and (\ref{vgd}), the only difference between the amplitude expressions of FP and FD pulses is a factor of $r/w$, i.e. $A_{\alpha }^{\text{(FD)}}=\frac{r}{w}A_{\alpha }^{\text{(FP)}}$, this factor is much slowly variant rather than the main term, thus the centroid of two envelopes of FP and FD pulses are basically same under a same order index $\alpha$. The next step is to solve the centroid surface of the envelope. For the fundamental ($\alpha=1$) FP and FD pulses, it is easy to solve the centroid trajectory as:
\begin{equation}
F(\mathbf{r},t)=z-ct+\frac{{{r}^{2}}}{2R}=0\label{as}
\end{equation}
because on this trajectory the envelope meets the maximum (the denominator term meets the minimum). However, for the higher-order ($\alpha>1$) cases, the profile of the envelope is more curved and the centroid surface would be deviated from the Eq.~(\ref{as}). Here we use the perturbation method to solve the centroid surface $F_\alpha(\mathbf{r},t)=0$ available for the higher-order pulse by:
\begin{equation}
F_\alpha(\mathbf{r},t)=z-ct+\frac{{{r}^{2}}}{2R}+d_\alpha(\mathbf{r})=0\label{ahs}
\end{equation}
where the perturbation displacement is given by Taylor series:
\begin{equation}
d_\alpha(\mathbf{r})={d_\alpha^\prime}\big\vert_{r^2=0} r^2+\frac{d_\alpha^{\prime\prime}\big\vert_{r^2=0}}{2}r^4+\cdots
\end{equation}
The differential in Taylor coefficient is to $r^2$ considering the even symmetry. The case of higher $\alpha$ requires more Taylor coefficients to retain for enough accuracy. Through numerical simulation, the retaining of the first term can correctly describe the location of pulse envelope for the cases of $\alpha$ up to 10 (see Supplementary Information~F). 

\begin{figure}[t!]
	\centering
	\includegraphics[width=0.9\linewidth]{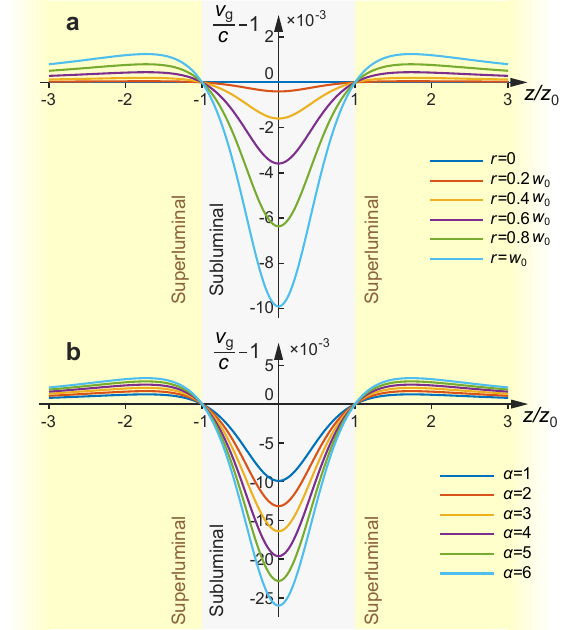}
	\caption{\textbf{Group velocity of HOSTP:} \textbf{a}, The group velocity distributions on various off-axis position from $r=0$ to $r=w_0$ versus propagation distance of the fundamental FP and FD pulses; \textbf{b}, The group velocity distributions on off-axis position $r=w_0$ of the HOSTP with various order indices $\alpha$ from 1 to 6. The gray and yellow regions mark the regions of subluminal and superluminal group velocity.} 
	\label{f4}
\end{figure}

Then, to solve the velocity of traveling of amplitude envelope, we make differential to the centroid surface equation $\delta F_\alpha(\mathbf{r},t)=0$ to the variables $z$ and $t$ and solve the group velocity as $v_\text{g}={\delta z}/{\delta t}$, the results are given by (see detailed derivation in Supplementary Information~F):
\begin{equation}
v_{\text{g}}^{(\alpha )}(r,z)=\frac{c}{1-\frac{\alpha +2}{6}\frac{{{z}^{2}}-z_{0}^{2}}{{{z}^{2}}+z_{0}^{2}}{{r}^{2}}}
\end{equation}
Figure~\ref{f4}\textbf{a} shows the group velocity distribution for the fundamental FP or FD pulse on various off-axis displacements from $r=0$ to $r=w_0$. For the on-axis ($r=0$) case, the group velocity shows a constant $c$, corresponding to the conventional speed of light in vacuum. While the off-axis displacement is increasing, the group velocity shows a distribution upon propagation distance. When the propagation is within the Rayleigh range, the group velocity is subluminal, and the traveling speed meets the maximum at focus and is decreasing to the constant $c$ with the propagation distance is up to Rayleigh length. When the propagation is beyond the Rayleigh range, the group velocity is superluminal, and the speed is firstly increasing and then decreasing approaching the limit of $c$ with the propagation distance from Rayleigh length to infinity. The larger the off-axis displacement the more serious the deviation from $c$ the group velocity distribution is. Such an abnormal distribution is induced by the spatiotemporal curved structure of pulse envelope upon propagation. The wavefront of HOSTP is approximatly flat when the propagation is at focus and at the infinity, and mostly curved at the position of Rayleigh length. We note that such abnormal distribution has be studied in prior structured pulse~\cite{lyons2018fast,saari2018reexamination}. Here we demonstrate that the similar distribution is similarly valid in our HOSTP and under our new definition of group velocity. The different group velocity distributions on $r=w_0$ for HOSTP with various orders $\alpha$ is shown in Figure~\ref{f4}\textbf{b}, where the subluminal and superluminal distribution is overall valid, and the higher order of the HOSTP off-axis displacement the deviated from $c$ the group velocity distribution is.

Another conclusion is that the value of group velocity is overall smaller than the phase velocity ($v_\text{p}/c-1$ is at $10^{-2}$ level while $v_\text{g}/c-1$ is at $10^{-3}$ level), in other words, the phase traveling is always faster than the amplitude envelope traveling, which interprets the phenomena of HOSTP that the cycle-switching is keeping forward upon propagation.

\vspace{0.2cm}
\noindent\textbf{Discussions} -- We propose and demonstrate a generalized family of focused few--cycle solutions to Maxwell's equations higher order spatiotemporal pulses (HOSTP) that extend the classical few--cycle FP and FD modes. We present local time amplitude-phase expressions for HOSTP; unlike conventional few--cycle pulses, both the amplitude and the phase are space--time nonseparable, enabling a broad range of novel spatiotemporal structures. The family exhibits clear higher--order properties analogous to spatial higher--order modes (more complex amplitude patterns and multiple Gouy--phase accumulations with increasing order), yet with genuinely spatiotemporal extensions. Notably, we observe a strongly nonseparable evolution in which a multi-cycle structure evolves into a single--cycle profile at focus, breaking the conventional ``fixed cycle--number'' description and opening directions for ultrafast structured light. 
Within the exact EDEPT construction, a stretch parameter \(p\), implemented through a spectral, phase mask that effects the term--wise shift \(\sigma\!\to\!\sigma-\Delta_m\)---stretches the temporal envelope while leaving the Gouy--phase coefficient unchanged (FD exceeds FP by one unit for the same \(\alpha\)). Consequently, temporal--revival locations governed by \(C_\alpha\tan^{-1}(z/z_0)\) are invariant with respect to \(p\), allowing pulse width to be tuned independently of modal order; this follows from the exact \(p\)-dependent FP/FD forms and the Gouy--coefficient analysis in the text and Supplementary Information.

To characterise such complex ST structure, we develop a unified approach to analysing the phase velocity and group velocity of structured pulses. The superluminal phase velocity accounts for the persistent forward cycle--switching, whereas the group velocity exhibits both subluminal and superluminal regions depending on transverse position and propagation distance, an expected feature of structured pulses that does not violate special relativity or causality. The same methodology applies to radial dynamics by taking differentials with respect to \(r\) and \(t\) in the isophase or centroid--surface equations; in most propagation scenarios, radial--velocity components are much smaller than the longitudinal component and are therefore omitted for brevity.

Because HOSTP introduce additional degrees of freedom for general ST waves, here we have illustrated spatiotemporal evolutions and topologies under representative conditions. Systematic links between broader parameter sets and the resulting dynamics, including tightly focused few, cycle pulses and vortex few, cycle pulses carrying orbital angular momentum, will be the focus of future studies.

Finally, the salient features of spatiotemporal evolution and propagation dynamics uncovered in HOSTP open avenues for fundamental studies (toroidal electrodynamics, topological optics, and light--matter interaction) and for practical applications such as precision metrology, particle acceleration, pulse compression, and communications.

\vspace{0.2cm}
\noindent\textbf{Acknowledgements} -- The authors acknowledge the supports of the MOE Singapore (MOE2016-T3-1-006), the UKs Engineering and Physical Sciences Research Council (grant EP/M009122/1, Funder Id: http://dx.doi.org/10.13039/501100000266), the European Research Council (Advanced grant FLEET-786851, Funder Id: http://dx.doi.org/10.13039/501100000781), and the Defense Advanced Research Projects Agency (DARPA) under the Nascent Light Matter Interactions program.

%{$^\dagger$zheludev@soton.ac.uk}
%$^*$shenyj15@tsinghua.org.cn

%\nocite{*}

\bibliography{apssamp}% Produces the bibliography via BibTeX.
%\\ \hspace*{\fill} \\
%
% ****** End of file apssamp.tex ******

\end{document}